\documentclass[twocolumn,preprintnumbers,amsmath,amssymb]{revtex4}

\usepackage{graphicx}
\usepackage{dcolumn}
\usepackage{bm,epsfig}

\begin{document}

\title{Physical properties and magnetic structure of the intermetallic CeCuBi$_{2}$ compound}

\author{C. Adriano$^{1}$}
\author{P. F. S. Rosa$^{1,2}$}
\author{C. B. R. Jesus$^{1}$}
 \author{J. R. L. Mardegan$^{1}$} \altaffiliation[Present address: ]{Deutsches Elektronen-Synchrotron DESY, Hamburg 22603, Germany} 
\author{T. M. Garitezi$^{1}$}
\author{T. Grant$^{2}$}
\author{Z. Fisk$^{2}$} 
\author{D. J. Garcia$^{3}$}
\author{A. P. Reyes$^{4}$} 
\author{P. L. Kuhns$^{4}$} 
\author{R. R. Urbano$^{1}$}
\author{ C. Giles$^{1}$}
\author{P. G. Pagliuso$^{1}$}

%\author{C. Adriano$^{1}$}
 %\author{P. F. S. Rosa$^{1,2}$}, C. B. R. Jesus$^{1}$,  J. R. L. Mardegan$^{1}$, \altaffiliation[Present address: ]{Deutsches Elektronen-Synchrotron DESY, Hamburg 22603, Germany} T. M. Garitezi$^{1}$, T. Grant$^{2}$, Z. Fisk$^{2}$, D. J. Garcia$^{3}$, A. P. Reyes$^{4}$, P. L. Kuhns$^{4}$, R. R. Urbano$^{1}$, C. Giles$^{1}$ and P. G. Pagliuso$^{1}$}

\affiliation{
$^{1}$Instituto de F\'isica \lq\lq Gleb Wataghin\rq\rq, UNICAMP, Campinas-SP, 13083-859, Brazil. \\
$^{2}$University of California, Irvine, California 92697-4574, U.S.A. \\
$^{3}$Instituto Balseiro, Centro Atomico Bariloche, CNEA and CONICET, 8400 Bariloche, Argentina.\\
$^{4}$National High Magnetic Field Laboratory, FSU, Tallahassee-FL, 32306-4005, U.S.A.}

\date{\today}

\begin{abstract}

In this work, we combine magnetization, pressure dependent electrical resistivity, heat-capacity, $^{63}$Cu Nuclear Magnetic Resonance (NMR) and X-ray resonant magnetic scattering experiments to investigate the physical properties of the intermetallic CeCuBi$_{2}$ compound. Our single crystals show an antiferromagnetic ordering at $T_{\rm N}\simeq$ 16 K and the magnetic properties indicate that this compound is an Ising antiferromagnet. In particular, the low temperature magnetization data revealed a spin-flop transition at $T$ = 5 K when magnetic fields of about $5.5$ T are applied along the $c$-axis. Moreover, the X-ray magnetic diffraction data below $T_{\rm N}$ revealed a commensurate antiferromagnetic structure with propagation wavevector $(0~0~\frac{1}{2}$) with the Ce$^{3+}$ moments oriented along the $c$-axis. Furthermore, our heat capacity, pressure dependent resistivity, and temperature dependent $^{63}$Cu NMR data suggest that CeCuBi$_{2}$ exhibits a weak heavy fermion behavior with strongly localized Ce$^{3+}$ 4$f$ electrons. We thus discuss a scenario in which both the anisotropic magnetic interactions between the Ce$^{3+}$ ions and the tetragonal crystalline electric field effects are taking into account in CeCuBi$_{2}$.

\end{abstract}

\pacs{71.20.Lp \sep 71.27.+a \sep 75.25.-j \sep 75.50.Ee}

%\keywords{}

\maketitle


\section{INTRODUCTION}

A series of rare-earth based intermetallic compounds is usually of great interest to explore the interplay between Ruderman-Kittel-Kasuya-Yoshida (RKKY) magnetic interaction, crystalline electrical field (CEF) and the Fermi surface effects frequently present in these materials. The Ce-based materials can have especially interesting physical properties that arise from the combination of these effects with a strong hybridization between the Ce$^{3+}$ 4$f$ and the conduction electrons. Therefore, these materials may present a variety of non-trivial ground states, including unconventional superconductivity (SC) and non-Fermi-liquid behavior frequently exhibited in the vicinity of a magnetically ordered state \cite{Review,Piers_JPCM_2001}.
Interestingly, some of these properties, such as the concomitant observation of unconventional SC and heavy fermion (HF) behavior, seem to be favored in systems with tetragonal structure. Well known examples are the Ce-based heavy fermions superconductors Ce$M$In$_5$ ($M$ = Co, Rh, Ir), Ce$_2M$In$_8$ ($M$ = Co, Rh, Pd), CePt$_2$In$_7$ ($M$ = Co, Rh, Ir, Pd) \cite{Thompson_Fisk_review115, NiniNP, pagliuso1, eric, curro} and CeCu$_2$Si$_2$ \cite{Steglich_CeCu2Si2, Stockert}.

%In this regard it is  to further investigate Ce-based intermetallic compounds in tetragonal structures and search for a realization of the above mentioned properties. 
Recent attention has been given to the Ce$T$X$_2$ family ($T$ = transition metal, $X$ = pnictogen) and in particular to the Ce$T$Sb$_2$ compounds \cite{CeTSb2}, which host ferromagnetic members with complex magnetic behavior, such as Ce(Ni,Ag)Sb$_2$. Their physical properties have motivated the investigation of the parent Ce$T$Bi$_2$ compounds \cite{CeTBi2, CeCuBi2_Acta, CeNiBi2_Takabatake}, although studies on the latter are rather rare. Thamizhavel \textit{et al.} \cite{CeTBi2} have shown that CeCuBi$_2$ orders antiferromagnetically with a N\'eel temperature of $T_{\rm N}$ = 11.3 K and an easy axis along the $c$-direction. Nevertheless, no detailed microscopic investigation regarding the relevant magnetic interactions have been presented so far. It is also intriguing that no HF superconductors have ever been discovered within the Ce$T$X$_2$ family. Another remarkable result is the breakdown of the De Gennes scaling revealed by non-Kondo members of the $RE$AgBi$_2$ \cite{ReAgBi2_Petrovic} and $RE$CuBi$_2$ \cite{ReCuBi2_Camilo} ($RE$ = rare earth) families. This usually indicates a complex and non-trivial competition between RKKY interactions and tetragonal CEF \cite{pagliuso2, Pagliuso_JAP2006, Serrano2}. 

In this work we report the physical properties and magnetic structure of CeCuBi$_2$ single crystals. CeCuBi$_2$ is an intermetallic compound that crystallizes in the tetragonal ZrCuSi$_2$-type structure (P4/nmm \cite{CeCuBi2_Acta} space group and lattice parameters $a$ = 4.555(4) $\rm{\AA}$ and $c$ = 9.777(8) $\rm{\AA}$) with a stacking arrangement of CeBi-Cu-CeBi-Bi layers. Our results revealed an antiferromagnetic ordering at $T_{\rm N}$ = 16 K, a higher value than reported previously \cite{CeTBi2, CeCuBi2_Acta} suggesting our crystals are of higher quality. In fact, we also found that the N\'eel temperature is suppressed in Cu-deficient crystals. For example, the compound CeCu$_{0.6}$Bi$_2$ orders antiferromagnetically at $T_{\rm N}$ = 12 K.
The magnetic structure determination of CeCuBi$_2$ revealed a propagation vector (0 0 $\frac{1}{2}$) with the magnetic moments aligned along the $c$-axis. A systematic analysis of the magnetization and specific heat data within the framework of a mean field theory with influence of anisotropic first-neighbors interaction and tetragonal CEF \cite{Pagliuso_JAP2006} allowed us to extract the CEF scheme for CeCuBi$_2$. It also led us to estimate the values of the anisotropic RKKY exchange parameters between the Ce$^{3+}$ ions.
In addition, the analyses of electrical resistivity under hydrostatic pressure and $^{63}$Cu Nuclear Magnetic Resonance (NMR) data suggest a scenario where CeCuBi$_2$ might display a weak heavy fermion behavior with rather strong localized Ce$^{3+}$ $4f$ electrons. 

\section{EXPERIMENTAL DETAILS}

Single crystals of CeCuBi$_{2}$ and LaCuBi$_2$ (a non-magnetic reference) were grown from Bi-flux, as reported previously \cite{ReCuBi2_Camilo}. The crystallographic structure was verified by X-ray powder diffraction and the crystal orientation was determined by the usual Laue method. The system was submitted to elemental analysis using a commercial Energy Dispersive Spectroscopy (EDS) microprobe and a commercial Wavelength Dispersive Spectroscopy (WDS). For oxygen free surface samples the stoichiometry is 1:1:2 with an error of 5\%. 

Magnetization measurements were performed using a commercial superconducting quantum interference device (SQUID). The specific heat was measured using a commercial small mass calorimeter that employs a quasi-adiabatic thermal relaxation technique. The in-plane electrical resistivity was obtained using a low-frequency ac resistance bridge and a four-contact configuration. Electrical-resistivity measurements under hydrostatic pressure were carried out in a clamp-type cell using Fluorinert as a pressure transmitting medium. Pressure was determined by measuring the superconducting critical temperature of a Pb sample.

X-ray resonant magnetic scattering (XRMS) measurements of CeCuBi$_2$ were carried out at the 4-ID-D beamline at the Advanced Photon Source of the Argonne National Laboratory-IL.  The sample was mounted on a cryostat installed in a four-circle diffractometer with the \textit{a}-axis parallel to the beam direction. This configuration allowed $\sigma$-polarized incident photons in the sample. The measurements were performed using polarization analysis, with a LiF(220) crystal analyzer, appropriate for the energy of Ce-L$_2$ absorption edge (6164 eV).

NMR experiments were performed at the National High Magnetic Field Laboratory (NHMFL) in Tallahassee-FL. A CeCuBi$_2$ single crystal was mounted on a low temperature NMR probe equipped with a goniometer, which allowed a fine alignment of the crystallographic axes with the external magnetic field. A silver wire NMR coil was used in this experiment. The field-swept $^{63}$Cu NMR spectra ($I$ = 3/2, $\gamma_N$/2$\pi$ = 11.285 MHz/T) were obtained by stepwise summing the Fourier transform of the spin-echo signal.

\section{RESULTS AND DISCUSSIONS}

\begin{figure}
\begin{center}
%\vspace{-1.1cm}
\includegraphics[width=1.0\columnwidth]{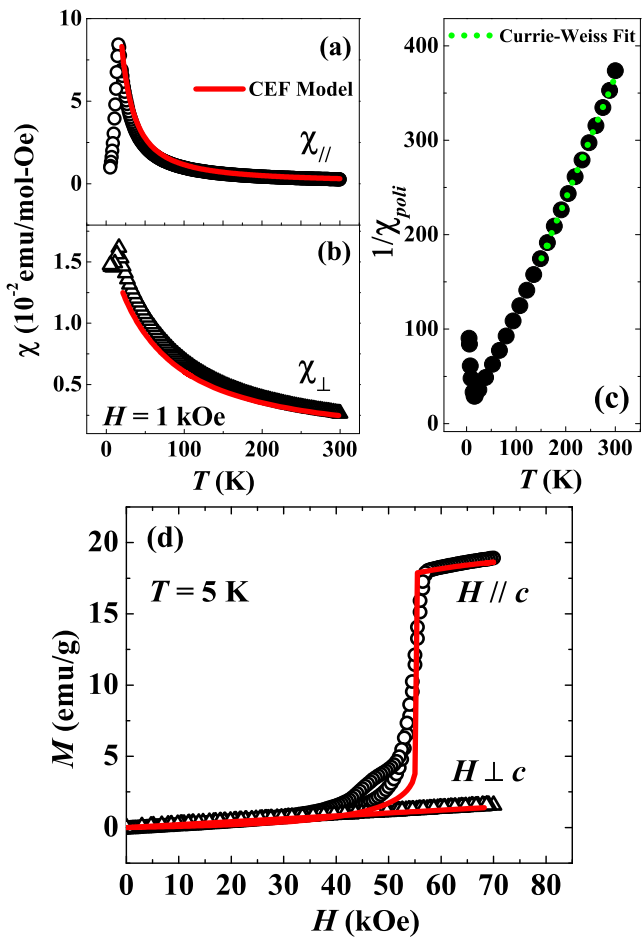}
\vspace{-0.7cm}
\end{center}
\caption{Temperature dependence of the magnetic susceptibility
measured with $H$ = 1 kOe applied (a) parallel
$\chi_{\parallel}$, and (b) perpendicular $\chi_{\bot}$ to the $c$-axis. (c) Inverse of the polycrystalline average 1/$\chi_{poly}(T)$. The green-dashed line represents a Curie-Weiss fit for $T >$ 150 K.
(d) Magnetization as a function of the applied magnetic field parallel (open circles) and perpendicular (open triangles) to the $c$-axis at $T$ = 5 K. The solid lines through the experimental points in Figs. 1a, 1b and 1d are best fits of the data using the CEF mean field model discussed in the text.} 
\label{fig:Fig1}
\end{figure}

Figures~\ref{fig:Fig1}a and 1b show the temperature dependence of the magnetic susceptibility $\chi(T)$ when the magnetic field ($H$ = 1 kOe) is applied parallel $\chi_{\parallel}$ (panel 1a) and perpendicular $\chi_{\bot}$ (panel 1b) to the crystallographic $c$-axis. These data show an antiferromagnetic (AFM) order at $T_{\rm N}\simeq$ 16 K and a low temperature magnetic anisotropy consistent with an easy axis along the $c$-direction. The ratio $\chi_{\parallel}/\chi_{\perp} \approx$ 4.5 at $T_{\rm N}$ is mainly determined by the tetragonal CEF splitting and reflects the low-$T$ Ce$^{3+}$ single ion anisotropy. The inverse of the polycrystalline 1/$\chi_{poly}(T)$ is presented in Fig.~\ref{fig:Fig1}c. A Curie-Weiss fit to this averaged data for $T>$ 150 K (dashed line) yields an effective magnetic moment $\mu_{eff}$ = 2.5(1) $\mu_{B}$ (in agreement with the theoretical value of $\mu_{eff}$ = 2.54 $\mu_{B}$ for Ce$^{3+}$) and a paramagnetic Curie-Weiss temperature $\theta_{p}$ = -23(1) K.

Figure~\ref{fig:Fig1}d displays the low temperature magnetization as a function of the applied magnetic field $M(H)$. The large magnetic anisotropy of CeCuBi$_2$ is also evident in these data.
We found an abrupt spin-flop transition from an antiferromagnetic to a ferromagnetic (FM) phase at $H\approx$ 55 kOe when the magnetic field is applied parallel to the $c$-axis (open circles) whilst a linear behavior is observed when the field is applied perpendicular to the $c$-axis (open triangles) for fields up to $H$ = 70 kOe.
Interestingly, the $M(H)$ data show a small hysteresis around $H\sim$ 50 kOe, suggesting a first order character for this field induced phase transition. The solid lines through the data points in Figs. 1a, 1b and 1d represent the best fits using a CEF mean field model discussed in detail ahead. 

\begin{figure}
\begin{center}
%\vspace{-1.0cm}
\includegraphics[width=0.95\columnwidth,keepaspectratio]{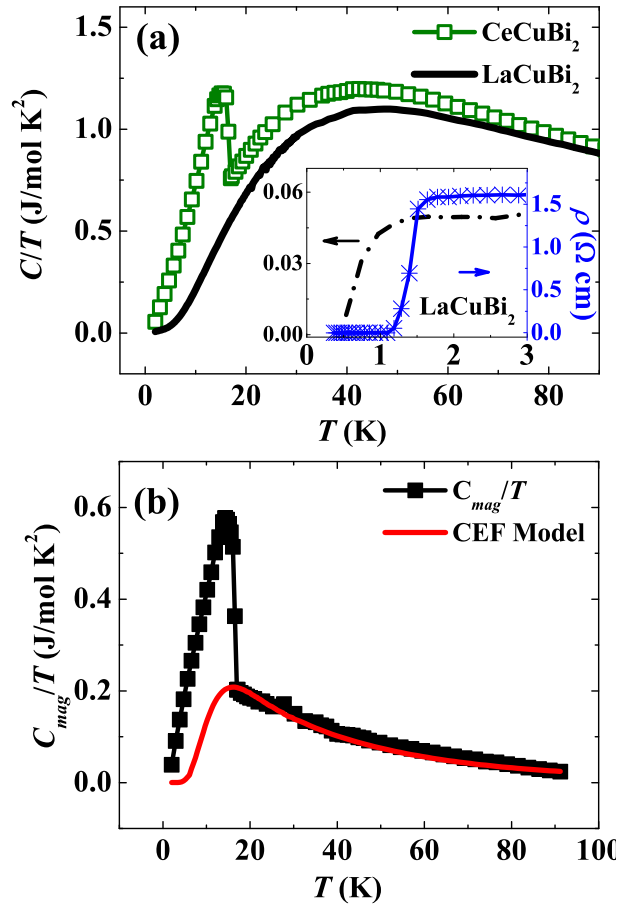}
\vspace{-0.7cm}
\end{center}
\caption{(a) $C(T)/T$ of CeCuBi$_2$ (open squares) and LaCuBi$_2$ (solid line) as a function of temperature. The inset shows the SC transition at $T\sim$ 1.3 K observed in the low-$T$ specific heat and electrical resistivity data of LaCuBi$_2$. (b) $C_{mag}(T)/T$ as a function of temperature. The solid line represents a Schottky-type anomaly resulted from the tetragonal CEF scheme.}
\label{fig:Fig2}
\end{figure}

The total specific heat divided by the temperature $C(T)/T$ as a function of temperature for CeCuBi$_2$ (open squares) is shown in Fig. 2a. The peak of $C(T)/T$ defines $T_{\rm N}$ = 16 K consistently with the AFM transition temperature observed in the magnetization measurements. Fig. 2b presents the magnetic specific heat $C_{mag}(T)/T$ of CeCuBi$_2$ (solid squares) after subtracting the lattice contribution from the non-magnetic reference LaCuBi$_2$ compound (solid line in Fig.~\ref{fig:Fig2}a).
The magnetic entropy recovered at $T_{\rm N}$ obtained by integrating $C_{mag}(T)/T$ in this temperature range (not shown) was found to be about $80\%$ of R$ln$2 (R $\sim$ 8.3 J/mol K). This suggests that the magnetic moments of the Ce$^{3+}$ CEF ground state are slightly compensated due to the Kondo effect. Although the presence of magnetic frustration and short range order  may also explain the magnetic entropy above $T_{\rm N}$. Yet from the $C_{mag}(T)/T$ data above $T_{\rm N}$, it is possible to estimate the Sommerfeld coefficient $\gamma$ by performing a simple entropy-balance construction [S($T_{\rm N}$ - $\epsilon$) = S($T_{\rm N}$ + $\epsilon$)] \cite{Hegger_PRL2000}. Thus, one obtains a $\gamma\sim$ 50-150 mJ/mole K$^2$, very consistent with the partly compensated magnetic moment of the CEF doublet at the transition.

The inset of Fig. 2a highlights the superconducting transition found for LaCuBi$_2$ at $T\sim$ 1.3 K. In fact, conventional superconductivity at similar temperatures has been previously reported for isostructural compounds of the La$M$Sb$_2$ family ($M$ = Ni, Cu, Pd and Ag) \cite{Muro}. The solid line in Fig. 2b represents a Schottky-type anomaly resulted from the tetragonal CEF scheme obtained from our analysis as discussed in the following.

In order to establish a plausible scenario for the magnetic properties of CeCuBi$_2$, we have analyzed the data presented in Figs. 1 and 2 using a mean field model including the anisotropic interaction from nearest-neighbors as well as the tetragonal CEF hamiltonian. For the complete description of the theoretical model, see reference \cite{Pagliuso_JAP2006}. This model was used to simultaneously fit $\chi(T)$, $M(H)$ and $C_{mag}(T)/T$ data for $T> 20$ K as a constrain. The best fits yield the CEF parameters: B$^{0}_{2}$ = -7.67 K, B$^{0}_{4}$ = 0.18 K and B$^{4}_{4}$ = 0.11 K; and two RKKY exchange parameters: $z_{AFM}*J_{AFM}$ = 1.12 K  and $z_{FM}*J_{FM}$ = -1.18 K, where $z_{AFM}$ = 2 ($z_{FM}$ = 4) is the Ce$^{3+}$ nearest neighbors with an AFM (FM) coupling, in this case, along the $c$-axis ($ab$-plane). The XRMS experiment suggests a magnetic structure compatible with this scenario, as will be discussed later on this work. It is worth emphasizing that the fits converged only when two distinct $J_{RKKY}$ exchange parameters were considered. Although CeCuBi$_2$ has an AFM ground state at zero field, the presence of FM fluctuations are evidenced by the presence of a spin-flop transition in the $M(H)$ data. The extracted parameters resulted in a CEF scheme with a $\Gamma^{(1)}_{7}$ ground state doublet $(0.99|\pm5/2\rangle - 0.06|\mp3/2\rangle)$, a first excited state $\Gamma^{(2)}_{7}$ $(0.06|\pm5/2\rangle + 0.99|\mp3/2\rangle)$ at 50 K and a second excited doublet $\Gamma_{6}$ $(|\pm 1/2\rangle)$ at 149 K.

The obtained CEF scheme and exchange constants account for the main features of the data shown in Figures 1 and 2, meaning that the magnetic anisotropy, the spin-flop transition and the Schottky anomaly in $C_{mag}(T)/T$ are all well explained by this model. However, it is important to notice that the CEF parameters obtained from the fits to macroscopic measurements data may not be as precise and unique. An accurate determination of the CEF scheme and its parameters does require a direct measurement by inelastic neutron scattering \cite{Christianson_CEF115}, while the mixed parameters of the wave functions may be compared with a X-Ray absorption study \cite{Severing_Ce115}.
Nonetheless, apart from a more precise determination of the CEF parameters, the analysis presented here suggests that the Ce$^{3+}$ 4$f$ electrons behave as localized magnetic moments. The only indication of a possible Kondo compensation is given by the partially recovered magnetic entropy at $T_{\rm N}$ ($\sim 80\%$ of R$ln$2) and by the rough estimate of $\gamma$.

Hence, in order to further investigate the presence of Kondo lattice behavior in CeCuBi$_2$ we have also performed pressure dependent electrical resistivity. Applied pressure is well known to favor the Kondo effect with respect to the RKKY interaction in Ce-based HF \cite{Review,Piers_JPCM_2001,Thompson_Fisk_review115}.

\begin{figure}
\begin{center}
%\vspace{-1.0cm}
\includegraphics[width=0.95\columnwidth,keepaspectratio]{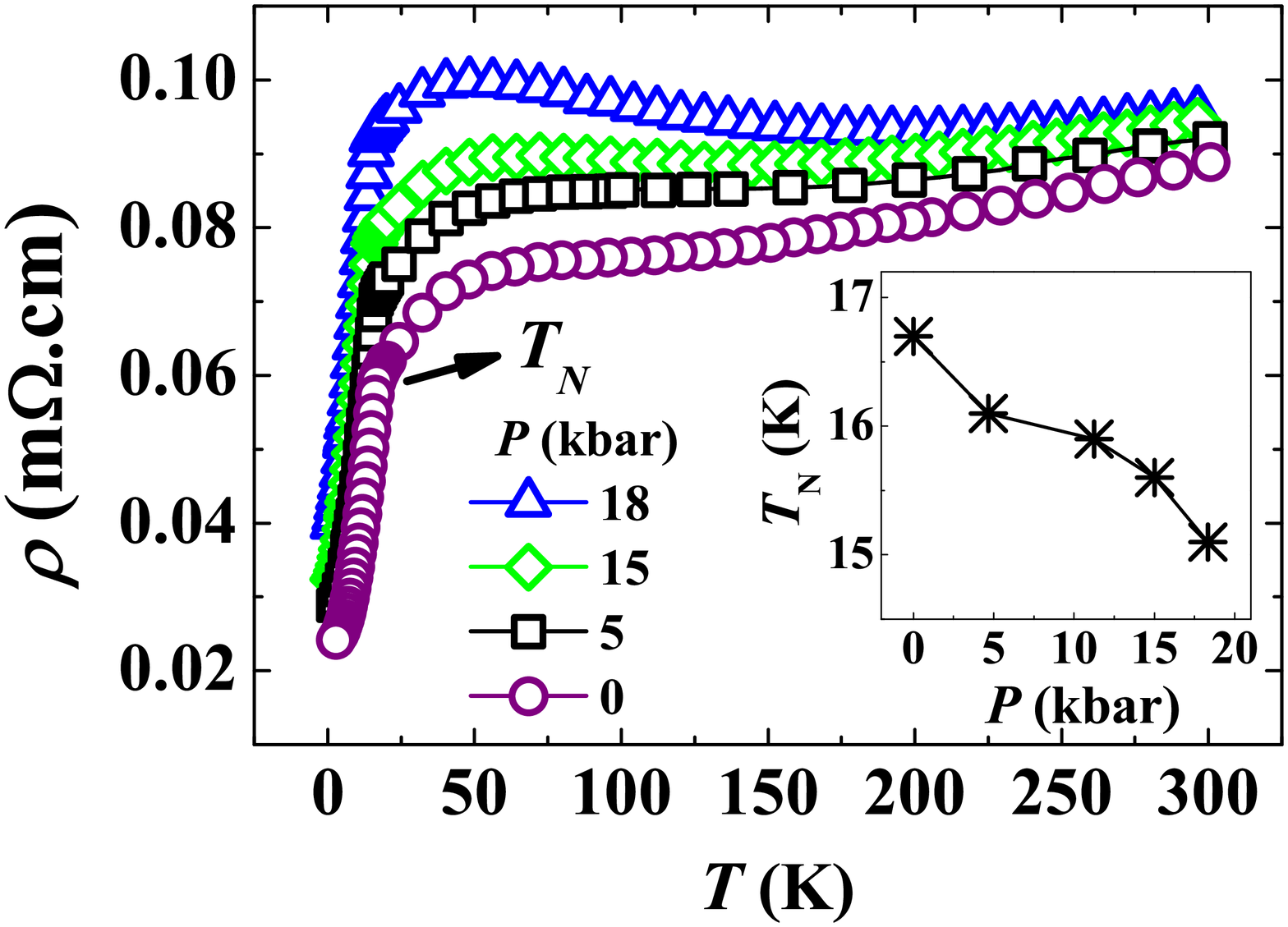}
\vspace{-0.9cm}
\end{center}
\caption{Temperature dependence of the electrical resistivity for different values of applied hydrostatic pressure up to 18 kbar. The inset shows the variation of $T_{\rm N}$ as a function of pressure.}
\label{fig:Fig3}
\end{figure}

The in-plane electrical resistivity $\rho(T,P)$ of CeCuBi$_2$ as a function of temperature for several pressures is summarized in Fig. 3. The electrical resistivity at ambient pressure first decreases with decreasing temperature, but it increases back for temperatures below $\sim$ 150 K. Then, $\rho(T,P=0)$ reaches a maximum at about 50 K and then drops abruptably after the magnetic scattering becomes coherent, as typically found for Ce-based HF \cite{Review,Piers_JPCM_2001,Thompson_Fisk_review115}.
At lower temperatures, a small kink is observed at $T_{\rm N}$ = 16 K. As pressure is increased, a small increase of the room-$T$ resistivity value is observed together with the decrease of $T_{\rm N}$. This effect can be seen in the inset of Fig. 3. Such suppression of $T_{\rm N}$ as a function of pressure is consistent with the increase of the Kondo effect on the Ce$^{3+}$ $f$-electrons. However, the slope d$T_{\rm N}$/d$P$ is relatively small and might be an indication that the Ce$^{3+}$ $f$-electrons remain rather localized in the studied pressure range.

To gain a more microscopic insight about the magnetic interactions present in CeCuBi$_2$, its magnetic structure was investigated by XRMS technique at the Ce-L$_2$ absorption edge in order to enhance the magnetic signal from Ce$^{3+}$ ions below $T_{\rm N}$. Magnetic peaks were observed in the dipolar resonant condition at temperatures below $\sim$ 16 K at reciprocal lattice points forbidden for charge scattering and consistent with a commensurate antiferromagnetic structure with propagation vector $(0~0~ \frac{1}{2})$.

To determine the possible irreducible magnetic representations $\Gamma^{\rm XRMS}$ associated with the space group $P4/nmm$, the propagation vector $(0~0~\frac{1}{2})$ and a magnetic moment at the Ce sites, we used the program SARA{\it h}~\cite{wills2000new}. The magnetic representation can be decomposed in terms of four non-zero irreducible representations (IRs - $\Gamma^{\rm XRMS}_{2}$, $\Gamma^{\rm XRMS}_{3}$, $\Gamma^{\rm XRMS}_{9}$ and $\Gamma^{\rm XRMS}_{10}$) written in Kovalev's notation~\cite{Kovalev_1993} . Within the possible IRs $\Gamma^{\rm XRMS}_{2}$ and $\Gamma^{\rm XRMS}_{3}$ correspond to a magnetic structure with the Ce magnetic moments pointing along $c$-direction and $\Gamma^{\rm XRMS}_{9}$ and $\Gamma^{\rm XRMS}_{10}$ correspond to Ce magnetic moments lyng in the $ab$-plane. Also, $\Gamma^{\rm XRMS}_{2}$ and $\Gamma^{\rm XRMS}_{10}$ correspond to a FM coupling of the Ce ions within the unit cell forming a  (+ + - -) sequence (model I), and $\Gamma^{\rm XRMS}_{3}$ and $\Gamma^{\rm XRMS}_{9}$ correspond to an AFM coupling of the Ce ions within the unit cell forming a (+ - - +) sequence (model II), both along the $c$-direction. 

Figure~\ref{fig:Fig4} shows typical results for one selected magnetic peak (0 0 5.5). Fig.~\ref{fig:Fig4}a  presents the resonant energy line shape showing a single peak 3 eV below the edge and compatible with a pure dipolar resonance. Fig.~\ref{fig:Fig4}b shows the intensity as a function of the angle $\theta$, where a pseudo Voigt fit shows a full width half maximum of 0.023$^o$. Fig. ~\ref{fig:Fig4}c presents the temperature dependence of the square root of the integrated intensity which is proportional to the magnetization of Ce$^{3+}$ ions. A pseudo Voigt peak shape was used to fit longitudinal $\theta$-2$\theta$ scans in order to obtain the integrated intensities and no hysteresis was observed by cycling the temperature.

The results presented in Figure~\ref{fig:Fig4} are consistent with a dipolar resonant magnetic scattering peak in which the magnetic intensity is found only in the $\sigma$-$\pi$' channel and disappears above T$_N$. This confirms the magnetic origin of the (0 0 5.5) reflection due to the existence of an AFM structure that doubles the chemical unit cell in the $c$-direction. For collinear magnetic structures, the intensity of the X-ray resonant magnetic scattering assumes a simple form for dipolar resonances\cite{Hill_Acta1996}:

\begin{figure}
\begin{center}
%\vspace{-1.2cm}
\includegraphics[width=0.9\columnwidth,keepaspectratio]{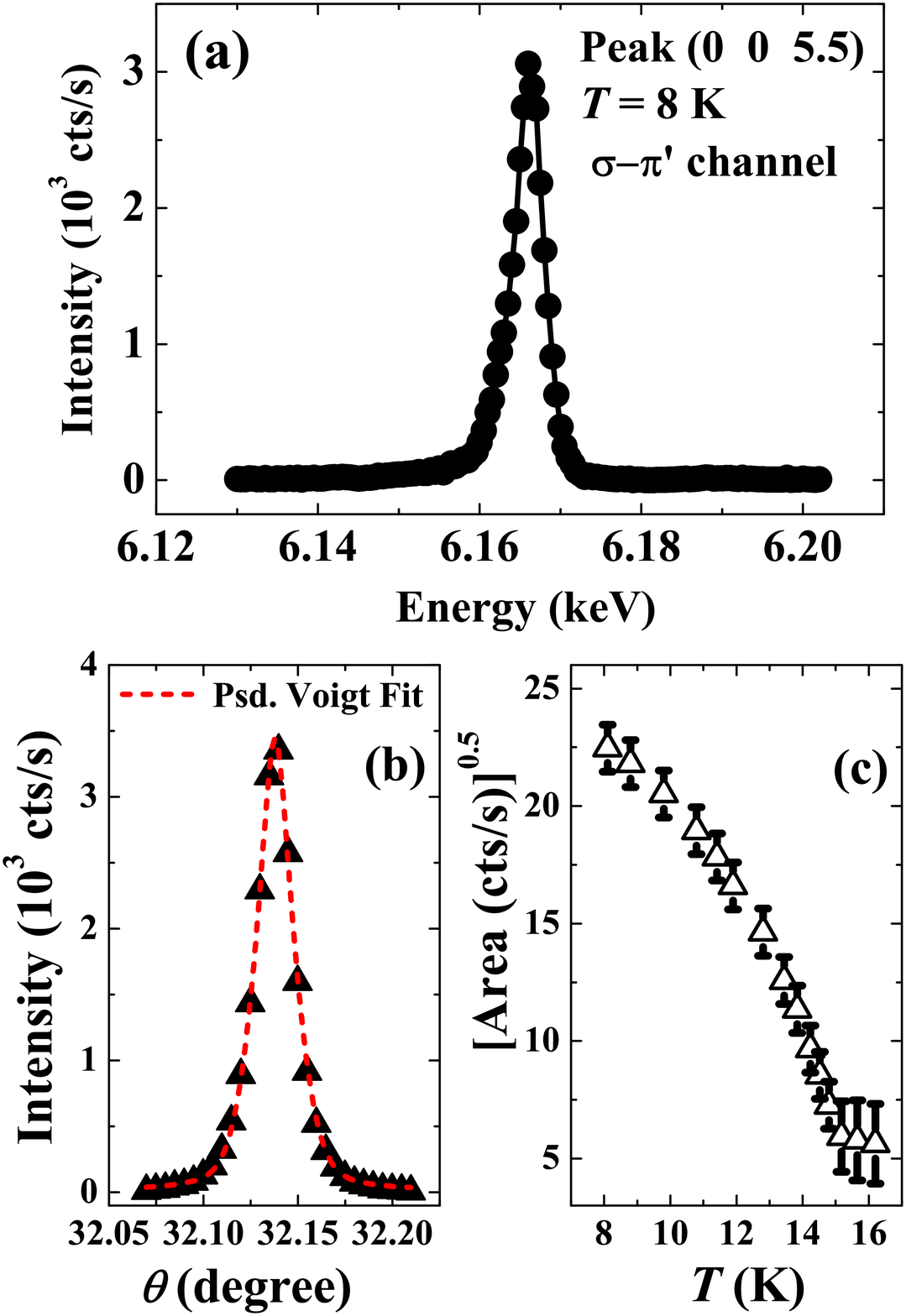}
\vspace{-1.0cm}
\end{center}
\caption{(a) Energy dependence of the XRMS signal of the Ce$^{3+}$ magnetic moment of CeCuBi$_2$.  (b) Intensity as a function of the angle $\theta$ of the crystal through the magnetic peak $(0~0~5.5)$ for $\sigma$-$\pi$' polarization
channel at the Ce $L_2$ absorption edge. (c) Square root of the intensity as a function of temperature measured with longitudinal ($\theta$-2$\theta$) from 8 to 16.5 K.}\label{fig:Fig4}
\end{figure}

\begin{eqnarray} \label{eq:equation1}
I\propto\frac{1}{\mu^{*}sin(2\theta)}\left|\sum_{n}\textit{f}_{n}^{E1}(\vec{k},\hat{\epsilon},\vec{k'},\hat{\epsilon'},\hat{z}_{n})e^{i\vec{Q}
\cdot \vec{R}_n}\right|^{2},
\end{eqnarray} where $\textit{f}_{n}^{E1}$ is the dipolar resonant magnetic form factor, $\mu^{*}$ is the absorption correction for asymmetric reflections, 2$\theta$ is the scattering angle, $\vec{Q}=\vec{k'}-\vec{k}$ is the wave-vector transfer, $\vec{k}$ and $\vec{k'}$ ($\hat{\epsilon}$ and $\hat{\epsilon'}$) are the incident and scattered wave (polarization) vectors, respectively. $\vec{R}_{n}$ is the position of the Ce \textit{n}th atom in the lattice, and $\hat{z}_{n}$ is the moment direction at the \textit{n}th site. The sum is over the \textit{n} resonant ions in the magnetic unit cell. 

\begin{figure}
\begin{center}
%\vspace{-1.2cm}
\includegraphics[width=1.0\columnwidth,keepaspectratio]{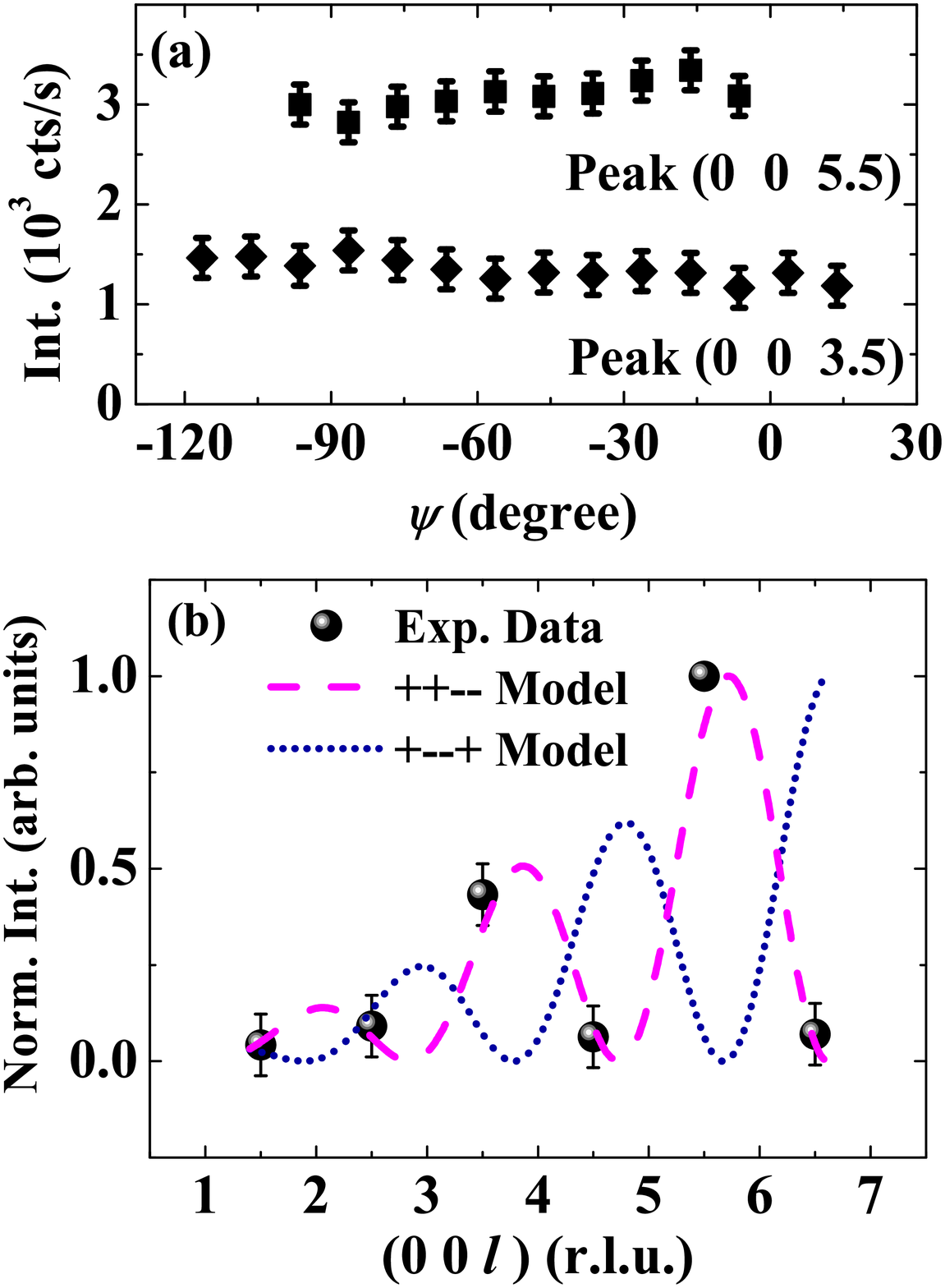}
\vspace{-1.2cm}
\end{center}
\caption{(a) Normalized intensity as a function of the azimuthal angle for (0 0 5.5) and (0 0 3.5) magnetic reflections.
(b) Experimental normalized intensity (solid circles) as a function of the \textit{l} at the
reciprocal space direction $(0,0,l)$ at 8 K for $\sigma$
- $\pi$' polarization channel at the Ce $L_2$ absorption
edge. The calculated intensities for two different magnetic couplings are presented as dashed and dot lines.} 
\label{fig:Fig5}
\end{figure}

The intensity variation of a magnetic peak as a function of the azimuthal angle can be used to determine the direction of the magnetic moment. In the case of a propagation vector (0 0 $\frac{1}{2}$) the azimuthal dependence of specular magnetic peaks will be constant if the moment is parallel to the c-axis and will show a sinusoidal dependence if the moment is perpendicular to the c-axis. Fig.~\ref{fig:Fig5}a shows the azimuthal dependence of the integrated intensity ($\sigma - \pi$' polarization channel) for two magnetic reflections (0 0 3.5) and (0 0 5.5). At each $\psi$ position a $\theta$ scan was measured and fitted using a Pseudo-Voigt function from which we extracted the integrated intensity value plotted in Fig.~\ref{fig:Fig5}a.

As we can see from data in Fig.~\ref{fig:Fig5}a, the azimuthal dependence of the integrated intensity of the magnetic peaks have a small variation (within error bars) and presents no sinusoidal periodicity. This result clearly indicates that the moment direction is parallel to the $c$-axis and is in good agreement with the susceptibility measurements of Fig.~\ref{fig:Fig1}a and b that also point for the $c$-axis as the easy magnetization axis. 

The magnetic coupling of the Ce atoms within the unit cell can be determined by comparing the experimental integrated intensity of several magnetic peaks with the calculated model (Eq.~\ref{eq:equation1}) \cite{Serrano_PRB74_2006,Cris_PRB2007,Cris_PRB2010}.

Simplifying the absolute square in Eq.~\ref{eq:equation1} for the reflections of the type $(0~0~\frac{l}{2})$, the magnetic intensity is proportional to $\sin^2(\theta +\alpha)* \cos^2$($2\pi l z) $ for the model (+ + - -) [or $\Gamma^{\rm XRMS}_{2}$] or $\cos^2(\theta +\alpha)*\cos^2$($2\pi l z)$ for the model (+ - - +) [or $\Gamma^{\rm XRMS}_{3}$], where $z$ is the position of Ce ions within the unit cell, $\theta$ is the Bragg angle and $\alpha$ is the angle between the vector $Q$ and the $c$-direction. Both magnetic representations consider the magnetic moments aligned parallel to $c$-direction.

Six magnetic peaks of the family $(0~0~\frac{l}{2})$ were measured at $T$ = 8 K and compared with the theoretical normalized intensities calculated using the model described by Eq. 1 (Fig. ~\ref{fig:Fig5}b). It is clear that the experimental data follows the (+ + - -) coupling. We can conclude that the correct magnetic structure corresponds to the $\Gamma^{\rm XRMS}_{2}$ irreducible representation, \textit{i.e.}, the magnetic moments are aligned parallel to $c$-axis with the (+ + - -) coupling. 

\begin{figure}
\begin{center}
%\vspace{-1.0cm}
\includegraphics[width=0.5\columnwidth,keepaspectratio]{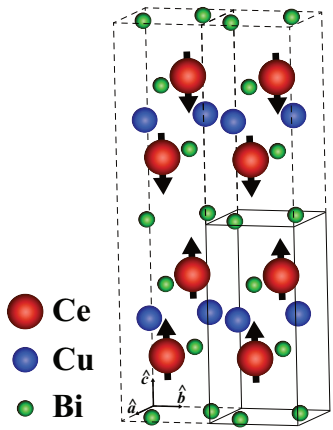}
\vspace{-0.5cm}
\end{center}
\caption{Schematic representation of the magnetic structure of CeCuBi$_2$. The dashed line defines two magnetic unit cells while the solid line bounds the chemical unit cells.}\label{fig:Fig6}
\end{figure}

\begin{figure}
\begin{center}
%\vspace{-1.0cm}
\includegraphics[width=1.0\columnwidth,keepaspectratio]{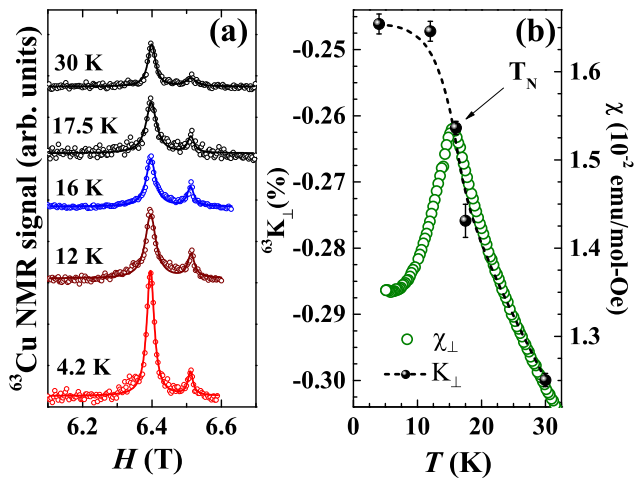}
\vspace{-0.5cm}
\end{center}
\caption{(a) $^{63}$Cu NMR spectra ($I$ = 3/2) for various temperatures around $T_{\rm N}$. (b) The corresponding temperature dependence of the Knight shift (left hand side scale) compared with the magnetic susceptibility at $H$ = 7 T (right hand side scale). The Knight shift was calculated as $^{63}K_{\bot} = \frac{(\nu/^{63}\gamma_N)- H_r}{H_r}$, with $\nu =$ 72 MHz and $H_r$ the peak position of the spectra for each temperature)}\label{fig:Fig7}
\end{figure}

Figure~\ref{fig:Fig6} represents the magnetic structure of CeCuBi$_2$ compound where we show two magnetic unit cells (dashed line) for better visualization of the spin coupling along the three directions. One chemical unit cell is represented by the solid line. The heretofore determined magnetic structure of CeCuBi$_2$ sheds some light on the global magnetic properties of this compound. The ferromagnetic coupling between the Ce$^{3+}$ moments in the plane is consistent with the presence of FM fluctuations which justifies the need to include two different exchange constants in our mean field model. Indeed, this proposed magnetic structure is compatible with the spin-flop transition to a ferromagnetic phase when a magnetic field is applied along the the $c$-axis.

Now, seeking for further microscopic information regarding the coupling of the Ce$^{3+}$4$f$ with the conduction electrons and/or neighboring atoms in CeCuBi$_2$, we have carried out temperature dependent $^{63}$Cu NMR measurements. NMR probes local interactions because it is site-specific and sensitive to both electronic charge distribution and magnetic spin.

Figure~\ref{fig:Fig7}a presents a few $^{63}$Cu NMR spectra ($I$ = 3/2) at temperatures around $T_{\rm N}$ with the magnetic field applied perpendicular to the $c$-axis.
Above $T_{\rm N}$ = 16 K, the $^{63}$Cu NMR spectra show a sharp single Lorentzian peak at $H\sim$ 6.4 T. We also observed a small peak at around 6.5 T which we associate with one of the $^{209}$Bi Zeeman split transitions ($I$ = 9/2). Although not the scope of the current investigation, further ongoing experiments will elucidate the origin of this signal. Also, for the low temperature spectra, below  $T_{\rm N}$, a weak and broad signal is observed at the $^{63}$Cu NMR line which might be associated with local field distribution at the $^{63}$Cu sites.

The Knight shift $^{63}K_{\bot}$ presented in Figure~\ref{fig:Fig7}b (left hand side axis) was obtained from Lorentzian fits of the main peaks shown in Figure~\ref{fig:Fig7}a. The $^{63}K_{\bot}$ is compared with the magnetic susceptibility $\chi_{\bot}$ measured with an applied magnetic field of 7 T. These data indicate that the Knight shift tracks the magnetic susceptibility down to $T_N\sim$ 16 K. Below this AFM transition, $^{63}K_{\bot}$ is driven by the internal field (hyperfine field) created by the Ce$^{3+}$ 4$f$ moments at the Cu sites. 

The Ce moments, slightly canted by the external field applied perpendicular to the $c$-axis create a weak ferromagnetic component in the plane responsible for the shift of the resonant peak towards lower fields. 
Moreover, the relatively small $^{63}K_{\bot}$ found for $^{63}$Cu NMR spectra in CeCuBi$_2$ compared to what is generally found in HF materials \cite{Yang_Urbano_PRL2009, Ohara_CeCu2Si2} is consistent with the weak hyperfine coupling constant estimated from the K-$\chi$ plot (not shown). This indicates that the Cu 3$d$ electrons are weakly hybridized with the Ce$^{3+}$ 4$f$ local moments \cite{DCox}. Within this scenario, the dipolar rather than the RKKY interaction seems to be the most relevant mechanism for the weak hyperfine coupling at the Cu sites. Additionally, the sign and strength of the coupling are not strongly influenced by the $c$-$f$ hybridization as expected in most heavy fermion materials. As such, one may speculate that the strong local moment character of the Ce$^{3+}$ 4$f$ magnetism in CeCuBi$_2$ is a dominant trend in Ce$T$X$_2$ family ($T$ = transition metal, $X$ = pnictogen) which makes these families less likely \cite{leticie} to host HF superconductivity, at least under ambient pressure.

Nevertheless, in a recent work \cite{Mizoguchi_CeNiBi2}, polycrystalline samples of CeNi$_{0.8}$Bi$_2$ have been reported as a heavy fermion superconductor with an AFM transition at $\sim$ 5 K and a SC transition at $\sim$ 4.2 K. The superconducting phase was claimed to be evoked by Ni deficiencies that would presumably create a different ground state than the one realized on single crystalline CeNiBi$_2$ \cite{CeTBi2}. The stoichiometric compound was earlier classified as a moderate HF antiferromagnet with $T_{\rm N}$ $\sim$ 5 K, and the presence of a zero resistance transition was associated to contamination of extrinsic Bi thin films. However, a more recent work has raised important questions about the intrinsic origin of the superconductivity in CeNi$_{0.8}$Bi$_2$. In the report, systematic studies on CeNi$_{1-x}$Bi$_2$ (with $1-x$ varying from 0.64 to 0.85) single crystals \cite{Lin_RNiBi2} revealed that the superconductivity in CeNi$_{0.8}$Bi$_2$ is more likely to be associated with the $T_c$ of the Bi thin films and/or secondary phases of the binaries NiBi and NiBi$_3$.

All the above arguments corroborate to our belief that the Ce$TX_2$ compounds do present strong local moment magnetism, with a moderate Kondo compensation implying a weak hybridization between the Ce$^{3+}$ 4$f$ ions and the conduction electrons. In absolute terms, this scenario does not favor a superconducting state.

\section{CONCLUSIONS}

In summary, we studied temperature dependent magnetic susceptibility, pressure dependent electrical resistivity, heat-capacity, $^{63}$Cu Nuclear Magnetic Resonance and X-ray magnetic scattering on CeCuBi$_{2}$ single crystals. Our data revealed that CeCuBi$_{2}$ orders antiferromagnetically at $T_{N}\simeq$ 16 K, a value higher than those previously reported for Cu-deficient samples. The detailed analysis of the macroscopic properties of CeCuBi$_{2}$ using a mean field model with a tetragonal CEF, enlightened by the microscopic experiments, allowed us to understand the magnetic anisotropy and the realization of a spin-flop first order-like transition in CeCuBi$_{2}$. These are very compatible with a magnetic field effect on the commensurate antiferromagnetic structure with propagation wavevector (0 0 $\frac{1}{2}$) and Ce moments oriented along the $c$-axis.
The combined analyses in this detailed investigation suggest that CeCuBi$_{2}$ presents a weak heavy fermion behavior with strongly localized Ce$^{3+}$ 4$f$ electrons subjected to dominant CEF effects and anisotropic RKKY interactions.

\begin{acknowledgments}
%\vspace{-1.5cm}
This work was supported by FAPESP (Grants No. 2009/09247-3, 2009/10264-0, 2011/01564-0, 2011/23650-5, 2011/19924-2, 2012/04870-7, 2012/05903-6 and 2013/20181-0), CNPq and CAPES-Brazil. The authors thank the 4-ID-D staff of the APS - ANL for the XRMS measurements. R.R.U. is grateful to Dr. Hironori Sakai for enlightening discussions. Work at NHMFL was performed under the auspices of the NSF through the Cooperative Agreement No. DMR-0654118 and the State of Florida. The authors acknowledge the Brazilian Nanotechnology National Laboratory LNNano for providing the equipment and technical support for the EDS
experiments.

\end{acknowledgments}

\bibliography{basename of .bib file}

\begin{thebibliography}{99}

\bibitem{Review} A. C. Hewson. \textit{The Kondo Problem To Heavy Fermions.} Cambridge University Press, Cambrige, 1993.


\bibitem{Piers_JPCM_2001} P. Coleman, C. P\'{e}pin, Q. Si, and R. Ramazashvili.  \textit{J. Phys. Condens. Matter} \textbf{R723-R738}, (2001).

\bibitem{Thompson_Fisk_review115} J. D. Thompson, and Z. Fisk. \textit{J. Phys. Soc. Japan} \textbf{81} 011002 (2012).

\bibitem{NiniNP}  S. Seo, Xin Lu, J-X. Zhu, R. R. Urbano, N. Curro, E. D. Bauer, V. A. Siderov, L. D. Pham, Tuson Park, Z. Fisk, and J. D. Thompson \textit{Nat. Phys.} \textbf{10}, 120 (2014).

\bibitem{pagliuso1} P. G. Pagliuso, C. Petrovic, R. Movshovich, D. Hall, M. F. Hundley, J. L. Sarrao, J. D. Thompson
and Z. Fisk,  Phys. Rev. B \textbf{64}, 100503(R)(2001).

\bibitem{curro} N. J. Curro, J. L.  Sarrao, J. D. Thompson, P. G. Pagliuso, S. Kos, A. Abanov and D. Pines, \textit{Phys. Rev. Lett.} \textbf{90}, 227202 (2003).

\bibitem{eric} E. D. Bauer, H. O. Lee, V. A. Sidorov, N. Kurita, K. Gofryk, J. X. Zhu, F. Ronning, R. Movshovich, J. D. Thompson and T. Park, \textit{Phys. Rev. B} \textbf{81}, 180507 (2010).

\bibitem{Steglich_CeCu2Si2} F. Steglich,  J. Aarts, C. D. Bredl, W. Lieke, D. Meshede, W. Franz, H. Schafer \textit{Phys. Rev. Lett.} \textbf{43} 1892 (1979).

\bibitem{Stockert} O. Stockert et al. Nature Phys. 7, 119-124 (2011).

 \bibitem{CeTSb2} A. Thamizhavel, \textit{et al} \textit{Phys. Rev. B} \textbf{68}, 054427 (2003).

\bibitem{CeCuBi2_Acta} J. Ye, Y. K. Huang, K. Kadowaki, T. Matsumoto,  \textit{Acta Cryst.} \textbf{C52}, 1323 (1996).

\bibitem{CeTBi2} A. Thamizhavel, \textit{et al} \textit{J. Phys. Soc. Japan} \textbf{72} 2632 (2003).

\bibitem{CeNiBi2_Takabatake} M. H. Jung, A. H. Lacerda, and T. Takabatake, \textit{Phys. Rev. B} \textbf{65}, 132405 (2002).

\bibitem{ReAgBi2_Petrovic} C. Petrovic, S. L. BudÕko, J. D. Strand, P. C. Canfield.  \textit{J. Mag. and Mag. Mat.} \textbf{261} 210 (2003).

\bibitem{ReCuBi2_Camilo} C. B. R. Jesus, M. M. Piva, P. F. S. Rosa, C. Adriano, and P. G. Pagliuso. \textit{J. Appl. Phys.} \textbf{115}, 17E115 (2014).

\bibitem{pagliuso2} P. G. Pagliuso, J. D. Thompson, M. F. Hundley, J. L. Sarrao, and Z. Fisk, Phys. Rev. B \textbf{63}, 054426 (2001).

\bibitem{Pagliuso_JAP2006} P. G. Pagliuso, D. J. Garcia, E. Miranda, E. Granado, R. Lora Serrano,
C. Giles, J. G. S. Duque, R. R. Urbano, C. Rettori, J. D. Thompson, M. F. Hundley and J. L. Sarrao,
J. Appl. Phys. \textbf{99}, 08P703 (2006).

\bibitem{Serrano2} R. Lora-Serrano, D. J. Garcia, E. Miranda, C. Adriano, C. Giles, J. G. S. Duque and P. G. Pagliuso, Phys. Rev. B \textbf{79}, 024422 (2009)

\bibitem{Christianson_CEF115} A. D. Christianson, E. D. Bauer, J. M. Lawrence, P. S. Riseborough, N. O. Moreno, P. G. Pagliuso, J. L. Sarrao, J. D. Thompson, E. A. Goremychkin, F. R. Trouw, M. P. Hehlen, and R. J. McQueeney. Phys. Rev. B \textbf{70}, 134505 (2004).

\bibitem{Severing_Ce115} T. Willers, Z. Hu, N. Hollmann, P. O. K\"orner, J. Gegner, T. Burnus, H. Fujiwara, A. Tanaka,
D. Schmitz, H. H. Hsieh, H.-J. Lin, C. T. Chen, E. D. Bauer, J. L. Sarrao, E. Goremychkin, M. Koza, L. H. Tjeng, and A. Severing. Phys. Rev. B \textbf{81}, 195114 (2010).

\bibitem{wills2000new} A. Wills, Physica B \textbf{276-278}, 680 (2000).

\bibitem{Kovalev_1993} O. V. Kovalev, {\it Representations of the Crystallographic Space Groups}, 2$^{nd}$ ed., edited by H. T. Stokes and D. M. Hatch (Gordon and Breach Science Publishers, Yverdon, Switzerland, 1993).

\bibitem{Hill_Acta1996} J. P. Hill and D. F. McMorrow, Acta Crystallogr., Sect. A: Found.
Crystallogr. \textbf{A52}, 236 (1996).

\bibitem{Hegger_PRL2000} H. Hegger, C. Petrovic, E. G. Moshopoulou, M. F. Hundley, J. L. Sarrao, Z. Fisk, and J. D. Thompson, Phys. Rev. Lett. \textbf{84} 4986 (2000).

\bibitem{Muro} Y. Muro, N. Takeda, M. Ishikawa, J. Alloys Compd \textbf{257} 23-29 (1997).

\bibitem{Serrano_PRB74_2006} R. Lora-Serrano, C. Giles, E. Granado, D. J. Garcia, E. Miranda, O. Ag\"uero, L. Mendon\c{c}a-Ferreira, J. G. S. Duque, and P. G. Pagliuso, Phys. Rev. B \textbf{74}, 214404 (2006).

\bibitem{Cris_PRB2007} C. Adriano, R. Lora-Serrano, C. Giles, F. de Bergevin, J. C. Lang, G. Srajer, C. Mazzoli,
L. Paolasini, and P. G. Pagliuso, Phys. Rev. B \textbf{76}, 104515
(2007).

\bibitem{Cris_PRB2010} C. Adriano, C. Giles, E. M. Bittar, L. N. Coelho, F. de Bergevin, C. Mazzoli, L. Paolasini, W. Ratcliff, R. Bindel, J. W. Lynn, Z. Fisk, and P. G. Pagliuso, Phys. Rev. B \textbf{81}, 245115 (2010).

\bibitem{Yang_Urbano_PRL2009} Yi-feng Yang, R. R. Urbano, N. J. Curro, D. Pines, E. D. Bauer, P. G. Pagliuso, \textit{Phys. Rev. Lett.} \textbf{103}, 197004 (2009) and references therein. 

\bibitem{Ohara_CeCu2Si2} Tetsuo Ohama, Hiroshi Yasuoka, D. Mandrus, Z. Fisk and J. L. Smith. \textit{J. Phys. Soc. Jpn} \textbf{64}, 2628 (1995). 

\bibitem{DCox} E. Kim, M. Makivic and D. L. Cox, Phys. Rev. Lett. \textbf{75}, 2015 (1995).

\bibitem{leticie} L. Mendon\c{c}a-Fereira, T. Park, V. Sidorov, M. Nicklas, E. M. Bittar, R. Lora-Serrano, E. N. Hering, S. M. Ramos, M. B. Fontes, E. Baggio-Saitovich, J. L. Sarrao, J. D. Thompson and P. G. Pagliuso, \textit{Phys. Rev. Lett.} \textbf{101}, 017005 (2008).

\bibitem{Mizoguchi_CeNiBi2} Hiroshi Mizoguchi, Satoru Matsuishi, Masahiro Hirano, Makoto Tachibana, Eiji Takayama-Muromachi, Hitoshi Kawaji, and Hideo Hosono, \textit{Phys. Rev. Lett.} \textbf{106}, 057002 (2011).

\bibitem{Lin_RNiBi2} Xiao Lin, Warren E. Straszheim, Sergey L. BudÕko, and Paul C. CanÞeld, \textit{J. Alloys Comp.} \textbf{554}, 304 (2012).

\end{thebibliography}

\end{document}